\documentclass[aps,prd,superscriptaddress,twocolumn,preprintnumbers,nofootinbib,10pt]{revtex4-2}
\usepackage{amsmath,amssymb}
\usepackage[dvipdf,dvips]{graphicx}
\usepackage{color}
\usepackage{hyperref}
\usepackage{url}
\usepackage{slashed}
\usepackage{subfigure}
\usepackage[usenames,dvipsnames]{xcolor}
\usepackage{amsmath}
\usepackage{amsfonts}
\usepackage{float} 
\usepackage{amssymb}
\usepackage{epsfig}
\usepackage{graphics}
\usepackage{euscript}
\usepackage{slashed}
\usepackage{epstopdf}
\usepackage[utf8]{inputenc}
\allowdisplaybreaks
\usepackage{tikz-feynman}
\usepackage{pifont}
\usepackage{dsfont}
\usepackage{bbold}
\usepackage[normalem]{ulem}
\usepackage{MnSymbol}
\usepackage{verbatim}
\usepackage{graphicx}
\usepackage{latexsym}

\def \and{\textmd{and}}

\def \be{\begin{equation}}
	\def \ee{\end{equation}}

\def \bea{\begin{eqnarray}}
	\def \eea{\end{eqnarray}}

\hypersetup{
colorlinks=true,
citecolor=blue,
citebordercolor=red,
linktoc=all,
linkcolor=blue,
urlcolor=blue
}

\usepackage{verbatim}
\usepackage{graphicx}
\usepackage{latexsym}

\newbox{\ORCIDicon}
\sbox{\ORCIDicon}{\large
                  \includegraphics[width=0.8em]{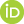}}

\begin{document}
	
	\title{Quark Propagator at one-loop in the Refined Gribov-Zwanziger framework }

	
	\author{Gustavo P. de Brito\,\href{https://orcid.org/0000-0003-2240-528X}{\usebox{\ORCIDicon}}} 
	\email{gp.brito@unesp.br}
	\affiliation{Departamento de Física, Universidade Estadual Paulista (Unesp), Campus Guaratinguetá, Av. Dr. Ariberto Pereira da Cunha, 333, Guaratinguetá, SP, Brazil}
	\author{Philipe De Fabritiis\,\href{https://orcid.org/0000-0001-5455-6889}{\usebox{\ORCIDicon}}} 
	\email{pdf321@cbpf.br} 
	\affiliation{CBPF $-$ Centro Brasileiro de Pesquisas Físicas, Rua Dr. Xavier Sigaud 150, 22290-180, Rio de Janeiro, Brazil}
	\author{Antonio D. Pereira\,\href{https://orcid.org/0000-0002-6952-2961}{\usebox{\ORCIDicon}}}
	\email{adpjunior@id.uff.br}
	\affiliation{Instituto de Física, Universidade Federal Fluminense, Campus da Praia Vermelha, Av. Litorânea s/n, 24210-346, Niterói, RJ, Brazil}

	
	\begin{abstract}
	The Refined Gribov-Zwanziger scenario is a local and renormalizable setup in which infinitesimal Gribov copies are eliminated and further non-perturbative effects are accounted for. The gluon propagator that arises from this framework fits lattice data very well in the Landau gauge. We investigate the coupling of quarks to this setting at one-loop order by computing the quark propagator. The fermionic sector is introduced by a minimal coupling and the non-perturbative effects are transmitted to the matter sector through gluonic loops which, in this case, carry information from the elimination of infinitesimal Gribov copies and the formation of condensates. We compare our findings with available lattice data both for the unquenched gluon propagator as well as for the quark propagator in the Landau gauge. Our results are comparable with those obtained in the Curci-Ferrari model at one-loop order. In particular, we are able to fit the unquenched gluon propagator and use the fixed parameters to predict the quark mass function and find good agreement with lattice data. However, the quark dressing function does not agree, even at a qualitative level, with lattice data in the infrared. This is agreement with the analogue computation in Curci-Ferrari model. Inspired by the developments in the Curci-Ferrari results, such a disagreement is likely to be cured by the inclusion of two-loops corrections. Finally, we compare the present minimal coupling with a non-perturbative matter coupling proposed in the Refined Gribov-Zwanziger literature.
	\end{abstract}

	\maketitle
	

	\section{Introduction}
		
	Yang-Mills (YM) theories play a fundamental role in our understanding of the microscopic structures of Nature. For instance, the Standard Model of particle physics (SM) is formulated in terms of YM theories. Yet YM theories present longstanding challenges. While their ultraviolet (UV) regime is well-understood due to asymptotic freedom \cite{Politzer:1973fx,Gross:1973id}, its infrared (IR) region is not properly understood due to its strongly correlated nature.	In particular, the understanding of the mechanism behind (color) confinement of  gluons is one of the main challenges connected with the IR sector/long-distances regime of YM theories~\cite{Greensite:2011zz,Brambilla:2014jmp}.
	
	The strong force in the SM is described by Quantum Chromodynamics (QCD), a YM theory with SU(3) gauge group coupled to quarks in the fundamental representation of the gauge group.  Remarkably, the presence of quarks does not spoil asymptotic freedom and being charged by the gauge group charge, i.e., being colored degrees of freedom, the property of color confinement carries over into the quark sector. This means that gluon and quarks are not seen as asymptotic states, i.e., as particles in the spectrum, but just appear as bound states.
		
	The fact that (color) confinement is present in the absence of (dynamical) quarks, often referred to as quenched QCD or simply pure YM theories, suggests that the gluonic sector might be the key object to describe such  mechanism. Nevertheless, the presence of quarks also triggers another challenging question to be addressed which is chiral-symmetry breaking. Despite of the fact that the quark sector does not have exact chiral symmetry,  it is often  said that it is approximate (at least for the light quarks). Running towards the IR, a growing mass is observed in lattice simulations, see, e.g., \cite{Oliveira:2018lln} and a first-principle mechanism to explain  chiral-symmetry breaking is still lacking. It is also often believed that color confinement and chiral-symmetry breaking might be intertwined and developing a consistent description of those phenomena solely based on the QCD Lagrangian is highly non-trivial since it requires the use and very good control of non-perturbative techniques. Different methods have been developed in order to describe confinement and chiral-symmetry breaking in an effective manner or from first-principles, see, e.g., \cite{Brambilla:2014jmp}.
	
	One particular avenue to be pursued in order to understand and describe quantitatively the IR sector of pure YM theories or QCD is to compute their correlation functions non-perturbatively. In the continuum, the computation of correlation functions requires the introduction of a gauge fixing which leads to the evaluation of non-gauge-invariant correlation functions. In principle they do not carry any physical information but as in any quantum field theory, they correspond to the building blocks of gauge-invariant correlation functions. Hence, to some extent, computing those correlation functions constitutes a first step towards the extraction of relevant physical information. Evaluating YM/QCD correlation functions non-perturbatively has turned into a solid research program that adopts many different approaches such as Dyson-Schwinger equations, Functional Renormalization Group equations, Variational methods, and others, see, e.g, \cite{Greensite:2011zz,Brambilla:2014jmp,vonSmekal:1997ohs,Alkofer:2000wg,Aguilar:2008xm,Alkofer:2008jy,Binosi:2009qm,Huber:2018ned,Pawlowski:2003hq,Fischer:2008uz,Dupuis:2020fhh,Cyrol:2016tym,Cyrol:2018xeq,Ihssen:2024miv,Siringo:2015wtx,Comitini:2024xjh}. 
	
	Moreover, effective models such as the Curci-Ferrari (CF) model have allowed for the use of perturbation theory together with specific renormalization schemes that lead to very accurate predictions of correlation functions in YM/QCD, see, e.g., \cite{Tissier:2010ts,Tissier:2011ey,Pelaez:2013cpa,Reinosa:2017qtf,Pelaez:2017bhh,Gracey:2019xom,Barrios:2020ubx,Pelaez:2020ups,Barrios:2021cks,Pelaez:2021tpq,Pelaez:2022rwx,Barrios:2024ixj}. See also \cite{Comitini:2023urc}. Notably, most of the development in this endeavor has focused in the Landau gauge due to its efficient numerical implementation and its algebraic simplifications on the continuum side.   
		
	The gauge-fixing procedure that is vastly used in the continuum is the so-called Faddeev-Popov (FP) method \cite{Faddeev:1967fc}. While the FP technique is well-defined and widely used in the perturbative regime (giving accurate predictions in the UV), the underlying assumptions of such a method are not well-grounded in the non-perturbative realm. As pointed out in \cite{Gribov:1977wm} and \cite{Singer:1978dk}, the FP method is ill-defined due to the existence of Gribov copies, i.e., different field configurations that satisfy the same gauge condition and can be connected by a gauge transformation. This is the so-called Gribov problem \cite{Vandersickel:2012tz}. 
	
	The elimination of Gribov copies requires a modification of the FP method in the IR and, therefore, can play an important role in the description of non-perturbative properties of non-Abelian gauge theories. It is important to emphasize that the existence of Gribov copies is not a particular pathology of a given gauge-fixing condition but lies on the non-trivial bundle structure of non-Abelian gauge theories as explored in \cite{Singer:1978dk}. For reviews on this issue, we refer to \cite{Vandersickel:2012tz,Sobreiro:2005ec,Vandersickel:2011zc}.
	
	In the context of the Landau gauge, one first-principle proposal was made in order to account for the existence of infinitesimal Gribov copies. It was initiated in \cite{Gribov:1977wm} and completed in \cite{Zwanziger:1989mf} leading to a local and perturbatively renormalizable action that modifies the FP gauge-fixed YM action in the Landau gauge. The action is known as the Gribov-Zwanziger (GZ) action and effectively removes infinitesimal Gribov copies from the path integral. The proposal hinges on restricting the path integral domain to a region where the FP operator is positive  that is known as the Gribov region. 
	
	The resulting action brings profound modifications with respect to the dynamics described by the FP action. The tree-level gluon propagator is suppressed in the IR while the FP ghost propagator is enhanced. The gluon propagator vanishes exactly at vanishing momentum. Such a property ensures positivity violation of the gluon propagator and this is often interpreted as a signal of confinement, since one cannot interpret gluons as physical particles in the spectrum. Moreover, the gluon propagator has a pair of complex poles. Similar behavior was found in different non-perturbative approaches such as Dyson-Schwinger equations and Functional Renormalization Group equations, see, e.g., \cite{vonSmekal:1997ohs,Alkofer:2000wg,Zwanziger:2001kw,Lerche:2002ep,Alkofer:2008jy,Huber:2018ned,Pawlowski:2003hq,Fischer:2004uk,Fischer:2006vf,Fischer:2008uz,Dupuis:2020fhh}. Such a combination of the behavior of gluon and ghost propagators in the IR characterizes what is known as the scaling solution for those correlation functions. 
	
	Nevertheless, simulations performed in larger lattices have revealed a surprising result \cite{Cucchieri:2007md,Cucchieri:2007rg,Sternbeck:2007ug,Cucchieri:2008fc,Bornyakov:2008yx,Bogolubsky:2009dc,Oliveira:2012eh}: The gluon propagator is finite but non-vanishing at zero momentum while the ghost propagator is not enhanced in the IR as in the previous case. This  established the so-called massive (decoupling) solution and it was also identified in functional methods, see, e.g., \cite{Aguilar:2008xm,Binosi:2009qm,Huber:2018ned,Fischer:2008uz}. In the GZ framework context, it was noticed in \cite{Dudal:2007cw,Dudal:2008sp} that the theory suffers from IR instabilities, i.e., leads to the formation of condensates. The inclusion of such condensates in the GZ action led to the so-called Refined Gribov-Zwanziger (RGZ) action which reproduces the qualitative features of larger lattice simulations \cite{Dudal:2010tf,Cucchieri:2011ig,Dudal:2018cli}. \\
	
	The RGZ action is local and renormalizable at all orders of perturbation theory, effectively removes infinitesimal Gribov copies, takes into account the aforementioned IR instabilities and describes the lattice data for the gluon and ghost propagator in the IR. In fact, the tree-level gluon propagator suffices to reproduce lattice data by a suitable fitting procedure \cite{Dudal:2010tf,Cucchieri:2011ig,Dudal:2018cli}. Actually, the RGZ action involves mass-parameters that are not free but fixed self-consistently through suitable gap equations \cite{Dudal:2011gd,Dudal:2019ing}. Unfortunately, their solutions up to date involve several approximations that make the corresponding predictions not accurate enough for quantitative comparison with lattice data. Hence, in the fitting procedure, those parameters are treated as free parameters. Clearly, a great open challenge in this framework is to compute those mass-parameters from a first-principle analysis and compare with the corresponding lattice fits. 
	
	One very important property of the (R)GZ framework is that it explicitly breaks BRST symmetry~\cite{Maggiore:1993wq,Baulieu:2008fy,Dudal:2009xh,Sorella:2009vt,Sorella:2010it,Capri:2010hb,Serreau:2012cg,Serreau:2013ila,Dudal:2012sb,Pereira:2013aza,Pereira:2014apa,Capri:2014bsa,Cucchieri:2014via,Schaden:2014bea,Schaden:2015uua}. Such a breaking is soft because it is proportional to the mass parameters that must vanish in the deep UV. Hence, the resulting action, in the UV, is the standard YM action quantized in the Landau gauge through the FP procedure and all the well-established properties of the BRST framework remains valid. Nevertheless, as one flows towards the IR, the mass parameters become  non-trivial and the (soft) BRST breaking takes place. In \cite{Capri:2015ixa} and subsequent works, see, e.g., \cite{Capri:2015nzw,Capri:2016aqq,Capri:2016gut,Capri:2017bfd,Capri:2018ijg}, a BRST-invariant reformulation of the RGZ action in the Landau gauge was proposed and their predictions for gauge and ghost fields correlation functions are completely identical. Such a construction has also enabled the exploration of the construction of the RGZ action in different gauges with respect to the Landau gauge, as linear covariant Curci-Ferrari and $R_\xi$-type gauges, see \cite{Capri:2017bfd,Pereira:2016fpn,Mintz:2018hhx}.
	
	The RGZ framework \cite{Dudal:2008sp} has achieved success in describing important aspects of YM theories in the IR. For instance, it was used to compute estimates for some glueball masses with reasonable agreement with lattice simulations, see \cite{Dudal:2010cd}. Moreover, it provides the correct sign for the (non-Abelian) Casimir energy in the so-called bag model, see \cite{Canfora:2013zna}. It has also been used to explore the phase diagram of YM-Higgs theories and its confined/non-confined regions, see \cite{Capri:2012ah,Capri:2013oja}. Nevertheless, the computation of correlation functions at higher orders in perturbation theory, even at one-loop order, in this framework is still in its early days. 
	
	The main obstacles for the limited amount of explicit computations lie in the complexity of the Feynman rules of the theory. The (R)GZ action contains many fields that render mixings and a large set of vertices which in turn proliferate very quickly the number of Feynman diagrams. Hence, simple one-loop calculations are, in some cases, very involved due to the large number of diagrams to be evaluated. Moreover, renormalizing within a suitable scheme to compare with lattice simulations is also a non-trivial task. Nevertheless, in recent years, progress in this direction has started to blossom. In \cite{Mintz:2017qri,Barrios:2024idr}, the ghost-gluon vertex was explored at one-loop order and compared with lattice simulations leading to good agreement and the gluon and FP ghost propagators were computed at one-loop order in \cite{deBrito:2024ffa}. In this work, the gluon propagator was fitted to the lattice data at fixed coupling and the result is quite encouraging. In \cite{deBrito:2023qfs}, the one-loop correction to colored scalar fields in the adjoint representation coupled to the RGZ action was computed and compared with available lattice data.
	
	Despite the promising results obtained within the RGZ framework, a major issue to be addressed is behind the RGZ-matter coupling. In principle, the whole modification to the FP procedure should produce direct modifications just on the pure-gauge sector and matter could be coupled in the usual minimal way as in standard YM-matter theories. Although this perspective seems to be the first attempt to be followed, different proposals on how to couple matter to the RGZ action were  pursued in \cite{Capri:2014bsa,Dudal:2013vha,Capri:2017abz}. Shortly, the motivations for a modified or non-perturbative matter coupling are twofold: On the one hand, minimally coupled matter fields to the RGZ do not suffer any influence on the propagator at tree-level. On the other hand, the original formulation of the RGZ action in the Landau gauge featured the soft-BRST breaking which was taken as a guiding principle to construct the non-perturbative coupling between matter and gauge fields, see \cite{Capri:2014bsa}. The proposals in \cite{Capri:2014bsa,Dudal:2013vha,Capri:2017abz} lead to a tree-level propagator for the matter fields which feature confining properties and compare well with lattice simulations. Yet such a proposal brings new mass parameters to the theory and allows for an independent fitting of gluon and matter field propagators. However, the introduction of new mass-parameters without the same geometrical status of those introduced in the RGZ in order to restrict the path integral to the Gribov region casts doubts if they are really necessary from a first-principle perspective. In fact, the confining nature of the gluon propagator could trigger a confining behavior for the matter-fields propagators indirectly through gluonic (and all the other complicated RGZ structure) loops. This was investigated in \cite{deBrito:2023qfs} for colored scalar fields and the results were promising in the sense that one-loop results could account for a reasonable fitting with lattice data without the ad hoc non-perturbative coupling between gauge fields and matter and without the introduction of new free (non-perturbative) parameters. 
	
	The main goal of this work is to investigate the quark propagator in the RGZ framework at one-loop order. More specifically, we couple quarks to the RGZ action be means of the minimal (or standard) procedure and evaluate the one-loop correction at one-loop order. To compare with lattice simulations, the unquenched gluon propagator is also evaluated at one-loop and a fitting with available lattice data \cite{Sternbeck:2012qs} is performed. This fit fixes the RGZ mass-parameters which are then used to predict the quark propagator (more specifically, the mass function) at one-loop order and our result is compared with \cite{Oliveira:2018lln}. We compare our findings with those obtained in the CF model~\cite{Pelaez:2014mxa,Barrios:2021cks}. Such a comparison is helpful in order to identify the limitations of the present results as well as perspectives that should be taken into account towards quantitative precision. We also refer the reader to works where the quark propagator was evaluated non-perturbatively by means of other methods see, e.g., \cite{Alkofer:2003jj,Fischer:2006ub,Alkofer:2008tt,Aguilar:2010cn,Aguilar:2012rz,Aguilar:2013hoa,Aguilar:2014lha,Braun:2014ata,Siringo:2016jrc,Cyrol:2017ewj,Aguilar:2018epe,Comitini:2021kxj,Gao:2021wun,Alkofer:2023lrl,Ihssen:2024miv,Fu:2025hcm} .

	This paper is organized as follows. We present a brief overview of the RGZ framework in Sec.~\ref{SecRGZ}. The inclusion of matter in the RGZ action is described in Sec.~\ref{SecThSetup}, where we define the main objects investigated here. We present and discuss our findings in Sec.~\ref{SecResults}, comparing them with previous results. Sec.~\ref{SecConclusions} contains our concluding remarks. Explicit analytical expressions for the quark propagator are collected in the appendix.
	
	\section{The Refined Gribov-Zwanziger framework in a nutshell}\label{SecRGZ}
	
	The action describing pure YM theories with gauge group $SU(N_c)$ in four-dimensional Euclidean space is
	\begin{align}
		S_{\rm YM} = \frac{1}{4} \int \! {\rm d}^4x~F_{\mu \nu}^a F_{\mu \nu}^a,    
	\end{align}
with the field strength tensor defined in terms of the gauge fields $A_\mu^a$ as $F_{\mu \nu}^a = \partial_\mu A_\nu^a - \partial_\nu A_\mu^a + g f^{abc} A_\mu^b A_\nu^c$. The symbol $f^{abc}$ represents the totally antisymmetric structure constants of $SU(N_c)$. The gauge coupling constant is written as $g$. According to our conventions, small case Latin indices, i.e.,  $a,b,c,\ldots$ range as $1,\ldots, N_c^2-1$.
	
	The functional integral quantization together with the FP procedure in the Landau gauge $(\partial_\mu A_\mu^a = 0)$ leads to the following gauge-fixed action:
	\begin{align}
		S_{\rm FP} = S_{\rm YM} + \int \! {\rm d}^4x \left(b^a \partial_\mu A_\mu^a + \bar{c}^a \partial_\mu D_\mu^{ab} c^b\right)\,,
		\label{Eq:SecRGZ.1}
	\end{align}
	with $D_\mu^{ab} = \delta^{ab} \partial_\mu - g f^{abc}A_\mu^c$ being the covariant derivative in the adjoint representation. The field $b^a$ is the Nakanishi-Lautrup field and it enforces the Landau gauge condition. The pair $\left(c^a, \bar{c}^a\right)$ are the FP ghosts. The functional integral for the gauge-fixed YM theories in the Landau gauge is given by
	\begin{align}
		\EuScript{Z} = \int \! [\EuScript{D}A] [\EuScript{D}c] [\EuScript{D}\bar{c}] [\EuScript{D}b] \, {\rm e}^{-S_{\rm FP}}.
		\label{Eq:SecRGZ.2}
	\end{align}
	
	The functional integral \eqref{Eq:SecRGZ.2} is the starting point for successful computations in the UV regime where perturbation theory is a reliable framework thanks to asymptotic freedom. This allows for a very accurate computational scheme where quantitative precision is achieved by the account of higher-order corrections in the perturbative series. However, due to the Gribov problem, it is well-known that the Landau gauge condition does not fix the gauge completely.   Since the FP procedure assumes that a given gauge condition picks up a single representative per gauge orbit, the existence of such spurious configurations demands an improvement of the FP procedure. In the Landau gauge, infinitesimal Gribov copies are related to the zero-modes of the FP operator $\mathcal{M}^{ab} \!\equiv\! -\partial_\mu D_\mu^{ab}$, see, e.g., \cite{Gribov:1977wm,Vandersickel:2012tz,Sobreiro:2005ec}. In \cite{Gribov:1977wm} Gribov  proposed a way to remove those (infinitesimal) copies, by restricting  the functional integral to a suitable set, the Gribov region $\Omega$, defined as 
	\begin{align}
		\Omega = \{ A_\mu^a \,\vert \, \partial_\mu A_\mu^a = 0 \,\, {\rm and} \,\, \mathcal{M}^{ab}>0 \}. 
		\label{Eq:SecRGZ.3}
	\end{align}
	Such a statement is mathematically meaningful because the FP operator in the Landau gauge is Hermitian and hence has a real spectrum. The Gribov region $\Omega$ has remarkable properties such as: It is bounded in every direction in field space and its boundary is known as the (first) Gribov horizon; It is convex and; All gauge orbits cross it at least once~\cite{DellAntonio:1991mms}. The last property ensures that the restriction of the path integral to $\Omega$ does not leave out any physical information. Nevertheless, it was proven later that there are still finite gauge copies inside $\Omega$, see \cite{vanBaal:1991zw}. To completely remove all Gribov copies, the path integral should be restricted to the so-called Fundamental Modular Region (FMR) which is, by definition, free of finite and infinitesimal gauge copies \cite{DellAntonio:1991mms}. However,  there is no  systematic way of implementing such a restriction of the path integral to the FMR up to date. Therefore, most of the progress in dealing with Gribov copies has been focused on the restriction of the path integral to $\Omega$. Albeit this is a partial solution to the Gribov problem it already corresponds to an improvement in the FP procedure and takes into account another assumption of such a method regarding the positivity of the FP operator. 
	
	The GZ action is an effective way of implementing the restriction of the functional integral to the Gribov region in a local and renormalizable framework, see \cite{Zwanziger:1989mf}. The path integral turns into
	\begin{align}
		\EuScript{Z}_{\rm GZ} = \int \! [\EuScript{D}\mu]_{\rm GZ} \, {\rm e}^{-S_{\rm GZ} + 4 \gamma^4 V (N^2_c-1)}.
		\label{Eq:SecRGZ.3.1}
	\end{align}
	The GZ action can be expressed as
	\begin{align}
		S_{\rm GZ} = S_{\rm FP} + S_{\rm H}\,,
		\label{Eq:SecRGZ.4}
	\end{align}
	with
	\begin{align}
		S_{\rm H} &= \int \! {\rm d}^4x \left[\bar{\varphi}^{ac}_\mu \mathcal{M}^{ab}(A) \varphi^{bc}_\mu - \bar{\omega}^{ac}_\mu \mathcal{M}^{ab}(A) \omega^{bc}_\mu \right. \nonumber \\
		&\left. + i g \gamma^2 f^{abc} A_\mu^a (\varphi_\mu^{bc} + \bar{\varphi}_\mu^{bc}) \right]\,.
		\label{Eq:SecRGZ.5}
	\end{align}
	The commuting $(\varphi, \bar{\varphi})$ and anti-commuting $(\omega, \bar{\omega})$ fields are auxiliary in the sense that they are introduced to localize the action. This means that one can integrate them out and end up with a non-local action which involves just the standard field content of gauge-fixed YM theories. Such a non-local term is also known as the Horizon function. For our purposes, the local version is what matters and hence we will deal with an extended field content. This means that the functional measure in \eqref{Eq:SecRGZ.3.1} is given by
	\begin{equation}
	[\EuScript{D}\mu]_{\rm GZ} \equiv [\EuScript{D}A][\EuScript{D}b][\EuScript{D}\bar{c}][\EuScript{D}c][\EuScript{D}\bar{\varphi}][\EuScript{D}\varphi][\EuScript{D} \bar{\omega}][\EuScript{D} \omega]\,.
	\label{Eq:SecRGZ.6}
	\end{equation}
	
	The parameter  $\gamma$ in Eq.\eqref{Eq:SecRGZ.5} has mass-dimension one and is the so-called Gribov parameter. It is not free but fixed by the gap equation
	\begin{align}
		\frac{\partial \mathcal{E}}{\partial \gamma^2} = 0\,,
		\label{Eq:SecRGZ.7}
	\end{align}
where $\mathcal{E}$ is the vacuum energy defined by $\EuScript{Z}_{\rm GZ} = {\rm e}^{- \mathcal{E} V}$ and $V$ is the Euclidean space volume. 
	Although the GZ action implements the restriction to the Gribov region in a local and renormalizable way, the gluon propagator vanishes at zero momentum. Such a property is exact and not a tree-level artifact. It also features an IR enhanced FP ghost propagator which is not observed by lattice simulations. Yet this framework has provided insightful understanding regarding the structure of correlation functions of elementary fields in non-perturbative YM theories and we refer the reader to \cite{Vandersickel:2012tz} for more details.
		
	The GZ action suffers from IR instabilities that must be taken into account as pointed out in \cite{Dudal:2007cw,Dudal:2008sp}. In particular, the dynamical formation of dimension-two condensates of gluons as well as of the auxiliary localizing fields must be included in the GZ framework \cite{Dudal:2011gd,Dudal:2019ing}. The effective setup in which those effects are included is the so-called RGZ action, 
	\begin{align}
		S_{\rm RGZ} = S_{\rm GZ} + S_{\rm c}\,,
		\label{Eq:SecRGZ.8}
	\end{align}
with $S_{\rm c}$ being the refining action that takes into account the effects of the formation of condensates. It reads
		\begin{align}
		S_{\rm c} = \int \! {\rm d}^4x \left[\frac{m^2}{2} A_\mu^a A_\mu^a + M^2 \left(\bar{\varphi}_\mu^{ab} \varphi_\mu^{ab} - \bar{\omega}_\mu^{ab} \omega_\mu^{ab}\right)\right]\,.
		\label{Eq:SecRGZ.9}
	\end{align}
	The mass parameters $(m,M)$ are also determined by their own gap equations \cite{Dudal:2011gd,Dudal:2019ing}. The inclusion of refining condensates modifies the gluon propagator in such a way that it fits very well the lattice data extracted from very large lattices, already at the tree-level in the IR \cite{Dudal:2010tf,Cucchieri:2011ig,Dudal:2018cli}. It also renders a ghost propagator that is not enhanced in the IR as in the case of the GZ action. The tree-level gluon propagator in the RGZ setup (in the Landau gauge) is
	\begin{align}
		\langle A_\mu^a(p) A_\nu^b(-p)\rangle_0 = \mathcal{D}_0(p) \, \delta^{a b} \, \Theta_{\mu \nu}, 
		\label{Eq:SecRGZ.10}
	\end{align}
with $\Theta_{\mu \nu} = \delta_{\mu \nu} - \frac{p_\mu p_\nu}{p^2}$ being the transverse projector in momentum space, and the form factor $D_0(p)$ is given by
	\begin{align}\label{GluonRGZtree}
		\mathcal{D}_0(p) = \frac{p^2 + M^2}{\left(p^2 + M^2\right) \left(p^2 + m^2\right) + \lambda^4},
	\end{align}
where we defined $\lambda^4 \equiv 2 \gamma^4 g^2 N_c$.
	It is worth mentioning that by expressing the gluon propagator in terms of $\lambda$ instead of $\gamma$, the loop expansion of the RGZ correlation functions matches the expansion in powers of $N_c \,g^2/(4\pi)^2$.
	As one can immediately verify, Eq.\eqref{GluonRGZtree} leads to a finite value for the gluon propagator at vanishing momentum. This is in agreement with the massive/decoupling behavior that is measured in recent lattice simulations \cite{Cucchieri:2007md,Cucchieri:2007rg,Sternbeck:2007ug,Cucchieri:2008fc,Bornyakov:2008yx,Bogolubsky:2009dc,Oliveira:2012eh}. Thus the RGZ framework yields a local and renormalizable action that implements the restriction of the path integral to the Gribov region in order to eliminate infinitesimal Gribov copies and takes into account the dynamical formation of condensates in the infrared regime. The modifications arising from the refinement of the GZ action provides a gluon propagator at tree level that is in good agreement with the most recent lattice data featuring positivity violation and complex poles \cite{Dudal:2010tf,Cucchieri:2011ig,Dudal:2018cli}. As observed in \cite{deBrito:2024ffa}, adding one-loop corrections to the gluon propagator does not affect such qualitative features leading to the conclusion that the leading radiative correction provides a first non-trivial hint towards the stability of the properties of the  gluon propagator in the IR.
	
	An important remark regarding the RGZ action is that, by turning off the Gribov parameter, i.e., setting $\gamma^2 = 0$ (thus setting $\lambda^2 = 0$), one ends up with an action where the auxiliary localizing fields can be exactly integrated out giving an unity. The resulting action is nothing but the action of the so-called CF model \cite{Tissier:2010ts,Tissier:2011ey}. Therefore, the CF model can be seen as a particular case of the RGZ action and for the sake of comparison of explicit computations, the one-loop corrections in the RGZ scenario should be compatible with those obtained in the simpler setup of the CF model when the Gribov parameter is taken to zero. 
	
	In the following, we will introduce fermionic matter minimally coupled to the gluons in the RGZ framework. The main goal is to investigate the infrared regime of the quark propagator in this scenario, as well as the modifications to the gluon propagator due to the addition of fermionic loops. The analysis will be performed at one-loop order in the same spirit of \cite{deBrito:2024ffa} and \cite{deBrito:2023qfs}. Comparing our findings with lattice data, we aim at providing hints for the confining nature of the quark propagator without the use of a non-perturbative coupling \cite{Capri:2014bsa,Dudal:2013vha,Capri:2017abz}. 
		
	\section{RGZ coupled to fermionic matter}\label{SecThSetup}

	Although the RGZ framework has been successful in the quenched limit, the inclusion of dynamical matter remains to be understood and investigated. In some works, the inclusion of matter in the context of the RGZ was discussed, see \cite{Capri:2014bsa,Dudal:2013vha,deBrito:2023qfs}. Particular focus was given to the Landau gauge due to its simplicity but  investigations are not limited to it, see \cite{Capri:2017abz}. Nevertheless, the gauge-matter coupling within the RGZ paradigm requires some fundamental explorations. A minimalist approach consists in including matter fields minimally coupled to the RGZ framework, i.e., adding the standard gauge-matter actions to the RGZ action. Since the tree-level matter sector does not change, one hopes that the modifications engendered by the elimination of (infinitesimal) Gribov copies will carry the relevant non-perturbative information from the infrared regime and influence the matter sector through quantum fluctuations. Such a perspective was studied in the case of RGZ-Scalar systems, where scalar fields, in the adjoint representation, were minimally coupled to the RGZ action in the Landau gauge, see \cite{deBrito:2023qfs}. By performing a one-loop computation of the scalar propagator, a quite reasonable agreement with available lattice data was achieved. In fact, this computation faced two major limitations: Firstly, the mass parameters of the RGZ framework were fixed by fitting the tree-level quenched gluon propagator with lattice data. The scalar propagator was fed by the values of those parameters and became a prediction of the framework. Hence, the results present a limitation in the sense that the fitting was performed within the quenched and hybrid approximation. Secondly, the computations were  crude with respect to the treatment of the gauge coupling, which was simply fixed. Although a parameterized analysis was performed, one clearly should properly address the runnning of the gauge-coupling in the presence of scalars and within the RGZ-Scalar framework. Yet the findings of \cite{deBrito:2023qfs} are encouraging in the sense that the minimal coupling between matter and the RGZ seems to be a good road to be taken to construct RGZ-matter models when quantum corrections are taken into account. 
	
	In the present work, we will consider the introduction of $N_f$ quarks in the fundamental representation of $SU(N_c)$ within the RGZ framework in the Landau gauge. The complete action is given by
	\begin{align}\label{RGZmatterAction}
		S = S_{{\rm RGZ}} + S_\psi,
	\end{align}
	with the quark sector described by
	\begin{align}
		S_\psi = \int \! {\rm d}^4 x \sum_{f=1}^{N_f}\left[\bar{\psi}^i_f \gamma^\mu D_\mu^{ij}\psi^j_f - m_\psi^f \bar{\psi}^i_f \psi^i_f    \right]\, .
	\end{align}
	The spinors $\psi^i_f$ are Dirac fermions in the fundamental representation of $SU(N_c)$ of flavor $f$ with mass $m_\psi^f$ and $\gamma^\mu$ are Dirac matrices in four-dimensional Euclidean space. The covariant derivative with $SU(N_c)$ generators $T_a^{ij}$ in the fundamental representation is expressed as
	\begin{align}
		D_\mu^{i j} = \delta^{ij} \partial_\mu - i g T_a^{i j} A_\mu^a\,,
	\end{align}
where lowercase Latin indices from the middle of the alphabet $i,j,\ldots$ correspond to indices of the fundamental representation of the gauge group.
	
\begin{figure}[t!]
\centering
	\includegraphics[width=0.35\textwidth]{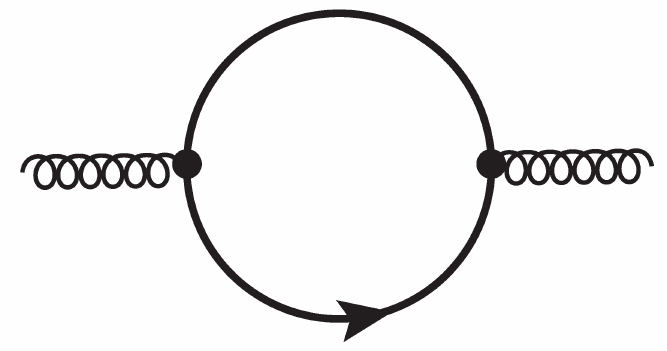}
	\caption{Quark contribution to the (unquenched) gluon propagator at one-loop order.}
	\label{quarkloop}
\end{figure}

	The main goal of this work is to compute the quark propagator at one-loop order in the RGZ framework minimally coupled to the quark sector and compare our findings with available lattice data. Besides being an important verification of the possibility of describing the quark sector in the IR with the RGZ-fermion model on its own, it is also another opportunity to verify if the minimal coupling in RGZ-matter models is a promising avenue to describe the IR behavior of matter fields propagators. However, including quarks minimally coupled to the gluons in the RGZ-fermion action, generates another vertex with respect to the RGZ, i.e., the quark-gluon vertex which affects the gauge sector by adding a new contribution/diagram to the gluon propagator at one-loop order, through the fermion-loop diagram represented in Fig.~\ref{quarkloop}. It is necessary to sum this diagram over each quark flavor with their corresponding masses to obtain the fully consistent contribution. Before proceeding with the computation, it is important to emphasize that the action \eqref{RGZmatterAction} is not BRST-invariant. In fact the RGZ sector breaks BRST explicitly but in a soft manner. However, as mentioned in the Introduction, a BRST-invariant RGZ framework is available and correlation functions of gluons, FP ghosts and quarks are equivalent in both formulations.

	The general form of the RGZ gluon propagator in the Landau gauge is given by
	\begin{align}
		\langle A_\mu^a(p) A_\nu^b(-p)\rangle =  \delta^{ab} \Theta_{\mu\nu}(p)\,\mathcal{D}(p^2).
	\end{align}
	The tree-level  expression of this propagator is given by e
	Eq.~\eqref{GluonRGZtree}. Recently, the RGZ gluon propagator was computed at one-loop order in the case of pure YM theories in \cite{deBrito:2024ffa}. We will rely on those results to calculate its unquenched version, by just adding the contribution arising from the quark sector. Since the fermion matter is minimally coupled to the gluon, the tree-level quark-gluon vertex is the standard one. In all explicit calculations, we employ dimensional regularization.
	
	The renormalization condition adopted in \cite{deBrito:2024ffa} for the computation of the gluon propagator at one-loop order fixes the corresponding two-point function to be equal to the tree-level one at a given renormalization scale $\mu$, that is,
	\begin{align}
		\langle A_\mu^a(p) A_\nu^b(-p)\rangle\bigg\vert_{p^2=\mu^2} = \langle A_\mu^a(p) A_\nu^b(-p)\rangle_0\bigg\vert_{p^2=\mu^2},
	\end{align}
	where the left-hand side denotes the one-loop RGZ gluon propagator and the right-hand side refers to the tree-level one (which is emphasized by the subscript $0$). The renormalization scale adopted there was $\mu = 1 \, {\rm GeV}$. The above renormalization condition is suitable to compare the analytical results with the lattice data, differently from the mass-independent $\overline{{\rm MS}}$-scheme adopted very often for explicit analytical computations in gauge theories. Moreover, this scheme allows for a direct comparison with the gluon propagator of the CF model in the infrared-safe scheme~\cite{Tissier:2011ey} if one takes the limit $\gamma \rightarrow 0$ to decouple the auxiliary fields, providing an important consistency check of the RGZ results.
	
\begin{figure}[t!]
\centering
		\includegraphics[width=0.4\textwidth]{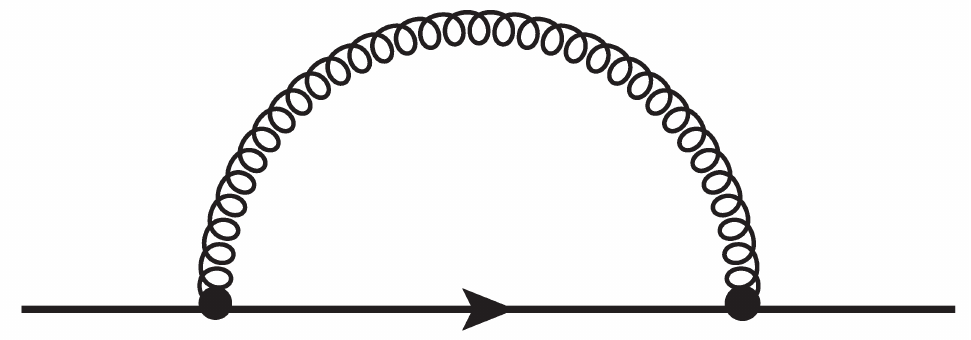}
	\caption{One-loop correction to the quark propagator.} 
	\label{quarkprop}
\end{figure}

	The one-particle irreducible (1PI) quark two-point function in the present RGZ-fermion model, for a given flavor, is
	\begin{equation}\label{key}
		\Gamma^{(2)}_{\psi \bar{\psi}}(p) = \left(-i \slashed{p} - m_\psi\right) + \delta\Gamma_{\psi \bar{\psi}}(p),
	\end{equation}
	where $\delta\Gamma_{\psi \bar{\psi}}(p)$ denotes the quantum corrections to the tree-level structure that can be written as $\delta\Gamma_{\psi \bar{\psi}}(p) = -i \slashed{p} \, \delta A(p) - m_\psi \, \delta B(p)$. The computation of the fermion propagator at one-loop order requires the explicit evaluation of just one diagram that arises from the quark-gluon vertex, see Fig.~\ref{quarkprop}.
	
	Putting together the tree-level and one-loop contributions lead to
	\begin{align}\label{1pi-2-point-quark}
		\Gamma^{(2)}_{\psi \bar{\psi}}(p) = -i \slashed{p} A(p) -m_\psi B(p),
	\end{align}
	where $A(p) \equiv 1 + \delta A(p)$ and $B(p) \equiv 1 + \delta B(p)$ are momentum-dependent structures that arise at one-loop order. Computing the Feynman diagram in Fig.~\ref{quarkprop} leads to the expressions of $\delta A$ and $\delta B$. This computation is simple enough and the analytical expression for it can be found in App.~\ref{AppAnalExp}. We remark that $\delta A(p)$ is finite at one-loop (it is even zero in the Yang-Mills limit where the mass-parameters are set to zero). Therefore, the only divergence is contained in the mass renormalization.
	
	The fermion propagator $S(p)$ is obtained by the inverse of the Eq.\eqref{1pi-2-point-quark} leading to
	\begin{align}
		S(p) = \mathcal{Z}(p) \left(\frac{i\slashed{p} - \mathcal{M}(p)}{p^2 + \mathcal{M}^2(p)}\right),
	\end{align}
	where we have defined the quark dressing function $\mathcal{Z}(p)$ and the quark mass function $\mathcal{M}(p)$ as
	\begin{align}
		\mathcal{Z}(p) &\equiv \frac{1}{A(p)} \quad {\rm and} \quad  \mathcal{M}(p) \equiv m_\psi \frac{B(p)}{A(p)}\,.
	\end{align}
	
	In the matter sector, we shall adopt a renormalization scheme that fixes the one-loop quark propagator to be the same as the one at tree-level at a given renormalization scale $\mu$, as it was done with the gluon propagator in \cite{deBrito:2024ffa}. This can be achieved by imposing renormalization conditions on the quark dressing function and the mass function at a given scale such that
	\begin{align}
		\mathcal{Z}(p=\mu) = 1 \quad {\rm and} \quad \mathcal{M}(p=\mu) = m_\psi\,. 
	\end{align}
	As remarked before, this renormalization scheme is suitable for comparing our analytical results with lattice data. Moreover, it allows us to compare our results with those of the CF model in the so-called IR-safe scheme, where equivalent renormalization conditions were imposed in the quark sector. Considering the limit $\lambda \rightarrow 0$, our analytical result reproduces exactly the expression presented in \cite{Pelaez:2014mxa}, showing that our findings are compatible with the CF results in the appropriate limit.
	
	The one-loop results for the gluon and quark propagators depend on the gauge coupling and the quark mass, but also on the RGZ mass-parameters $\{\lambda, m, M\}$, that can be fixed, in principle, by solving their corresponding gap equations self-consistently. Although this would be the ideal approach in order to test the consistency of the RGZ framework against lattice data this corresponds to a highly non-trivial task which goes far beyond the scope of the present work. As described before, the RGZ mass-parameters will be fixed by fitting the unquenched gluon propagator with available lattice data from \cite{Sternbeck:2012qs}.
	
	The renormalization group (RG) evolution of the gauge coupling plays an important role in the quark propagator, as we will see in the next section. However, unfortunately, the RG analysis in the context of the RGZ framework is still very scarce. The situation becomes worse for RGZ-matter systems. We partially amend this problem by using a toy model for the RG flow of the gauge coupling, inspired by the one-loop beta-function of YM theory in the minimal subtraction scheme, as in \cite{Barrios:2024idr},
	\begin{align}\label{RunningCoupling}
		g^2(p) = \frac{g_0^2}{1 + \frac{g_0^2}{16 \pi^2} \left(\frac{11}{3}N_c - \frac{2}{3}N_f\right) \, \log\left(\frac{p^2 + \Lambda^2}{\Lambda^2}\right)}\,,
	\end{align}
	where $\Lambda$ stands for an IR regulator that avoids the perturbative Landau pole and $g_0^2 = g^2(p=0)$. Taking the limit $\Lambda \rightarrow \infty$ we recover the case of a constant (fixed) gauge coupling $g^2_0$. For concreteness, we  choose  $\Lambda$ to be sitting  at the would-be Landau pole of the  standard running in YM theory, that is, $\Lambda = \mu \exp\left[-\frac{1}{2  \beta_0 g_0^2}\right]$, where we have defined $\beta_0 = \frac{1}{16\pi^2} \left(\frac{11}{3}N_c - \frac{2}{3}N_f\right)$.
	
	The ghost propagator in the unquenched case, at one-loop order, is the same as in the pure RGZ limit since a quark-ghost vertex does not exist. However, the ghost-two-point function can receive indirect contributions from the quark sector through RG effects affecting the running of the gauge coupling, for instance. We do not take this into account in this work.
	
	We have constructed the setup in which we will carry on our investigations. In the following, we will present the one-loop results for the QCD-like propagators, comparing them with available lattice data and discussing their implications, commenting also on previous findings using other approaches. In the following, we will restrict ourselves to the case $N_c = 3$ and $N_f = 2$ with degenerate masses to compare with available lattice data \cite{Oliveira:2018lln}.
	
	\section{Results and discussion}\label{SecResults}

	The strategy adopted in this work to compare our analytical results with lattice data in the following. The gluon propagator  is more sensitive to the mass-parameters $\{\lambda, m, M\}$, which are defined in the gauge sector and are already present at leading order in the gluon propagator the quark sector indirectly through gluonic  loops. Furthermore, all the quantities studied here have a  significant sensitivity to variations of the gauge coupling, and the gluon propagator is not strongly affected by different choices of  the quark mass. Therefore, it would be natural to fit the mass-parameters $\{\lambda, m, M\}$ using the unquenched gluon propagator lattice data, analyzing different values for the gauge coupling, but adopting a fixed value for the quark mass $m_\psi$, chosen  to be the lattice value for the quark mass function taken at a given renormalization point. Then, we can use the fitted parameters as an input for studying the functions $\mathcal{M}(p)$ and $\mathcal{Z}(p)$. Naturally, this strategy privileges the gluon propagator, guaranteeing the quality of its result since we use this data to fix the mass-parameters, and allow us to investigate the outcome of the minimal coupling prescription in the quark sector. This  seems to be a reasonable pathway since the RGZ framework is built from the gluon sector in the first place, and one of our goals is precisely to analyze the underlying repercussion of the removal of Gribov copies to the coupling of matter fields.
	
	More specifically, let us describe the methodology adopted in this work. First of all, we fix the gauge coupling at vanishing momentum, $g_0$,  and employ the toy running for the gauge coupling defined in Eq.\eqref{RunningCoupling}. We systematically  repeat the procedure described below for different values of $g_0$ ranging from $0$ to $9$ with $0.05$-sized steps. We also consider different values for the renormalization point, taken as $\mu = 1, 2, 3, 4 \, {\rm GeV}$.  The quark mass $m_\psi$ is fixed by the lattice value of the quark mass function at  the corresponding scale, i.e., $ m_\psi = \mathcal{M}_{\rm Lt}(p=\mu)$. Having the gauge coupling, the renormalization point, and the quark mass as inputs, we use the analytical expression for the unquenched gluon propagator form factor $\mathcal{D}(p)$ to fit the  mass-parameters $\{\lambda, m, M\}$. A global multiplicative factor is necessary to account for the lattice data normalization. The lattice data used is the one reported in~\cite{Sternbeck:2012qs}. Then, for the given values of $g_0$ and $\mu$, we use the mass-parameters obtained from the fitting of the unquenched gluon propagator to obtain $\mathcal{M}(p)$. We are thus ready to compare our analytical result with the quark propagator lattice data of \cite{Oliveira:2018lln}. It is worth pointing out that at this stage all parameters are fixed and thus, for given values of $g_0$ and $\mu$,  the mass function becomes a prediction within the described setup. Furthermore, we also compare our analytical results for $\mathcal{Z}(p)$ with available lattice data \cite{Oliveira:2018lln}, adjusting only a global multiplicative factor, with a standard fitting routine of \textsc{Mathematica}. Finally, we evaluate the error associated with the comparison between our result and the lattice data for $\mathcal{M}(p)$ using a $\chi^2$-type test,\footnote{Here we are using $\chi^2_M \equiv \sum_i \frac{\left[M_{\rm Th}(p_i) - M_{\rm Lt}(p_i)\right]^2}{M_{Lt}(p_i)}$, where $M_{\rm Th}$ and $M_{\rm Lt}$ refer to the analytical and lattice results, respectively.} for different coupling values, selecting the one that minimizes $\chi^2$. 
	
	We stress that the error associated with the gluon propagator data is negligible in comparison with the one associated with the quark mass function. This is reasonable since we mass-parameters. Moreover, the error associated with $\mathcal{Z}(p)$ is not being considered in this analysis because lattice data is poorly described by our analytical result as we will see later on. This is very similar, at least qualitatively, with the results obtained in the CF model in \cite{Pelaez:2014mxa}. At this stage a comment is in order: The perturbative scheme that we have employed in this work are not able to capture the effects of chiral-symmetry breaking. This means that, the closer we get to the chiral limit, the worse our results will be. For this reason, we decided to use lattice data associated with $M_\pi = 422 {\rm MeV}$ in order to keep us as far as possible from the chiral limit. A proper account of dynamical chiral-symmetry breaking effects deserves a thorough investigation.
	
\begin{figure}[t!]
		\begin{minipage}[b]{1.0\linewidth}
			\includegraphics[width=\textwidth]{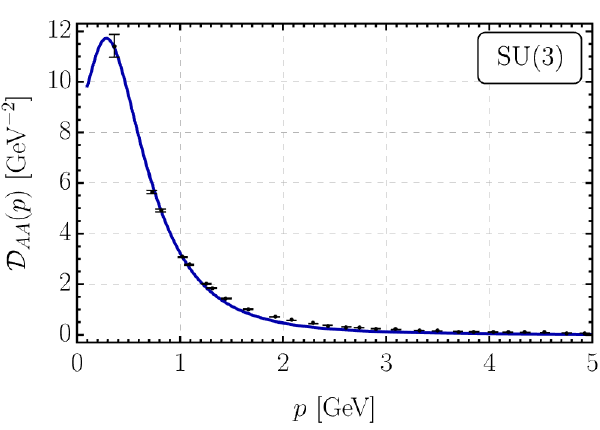}
		\end{minipage} \hfill
		\caption{Gluon propagator $\mathcal{D}(p)$  with $g_0 = 7$, $m_\psi = 37 \, {\rm MeV}$, $\lambda^2 = 2.817$, $m^2 = -0.935$, $M^2 = 7.299$, and $\mathcal{A_D} = 0.747$. Lattice data \cite{Sternbeck:2012qs} for $SU(3)$ with $N_f=2$ and $M_\pi = 422 \, {\rm MeV}$.}
		\label{Dpsifit}
	\end{figure}

\begin{figure}[t!]
		\begin{minipage}[b]{1.0\linewidth}
			\includegraphics[width=\textwidth]{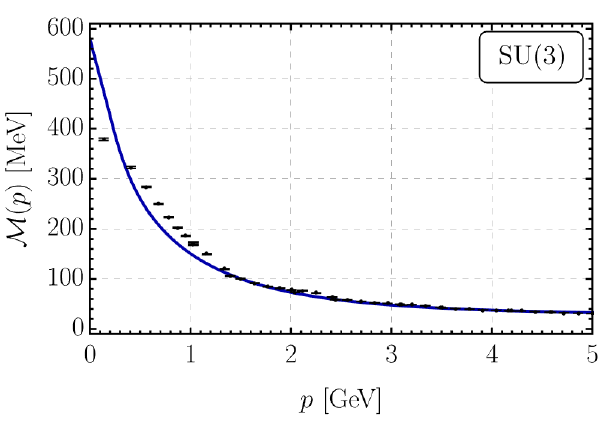}
		\end{minipage} \hfill
		\caption{Quark mass function $\mathcal{M}(p)$ with $g_0 = 7$, $m_\psi = 37 \, {\rm MeV}$, $\lambda^2 = 2.817$, $m^2 = -0.935$, $M^2 = 7.299$. Lattice data \cite{Oliveira:2018lln} for $SU(3)$ with $N_f=2$ and $M_\pi = 422 \, {\rm MeV}$.}
		\label{Mpsifit}
	\end{figure}

\begin{figure}[t!]
		\begin{minipage}[b]{1.0\linewidth}
			\includegraphics[width=\textwidth]{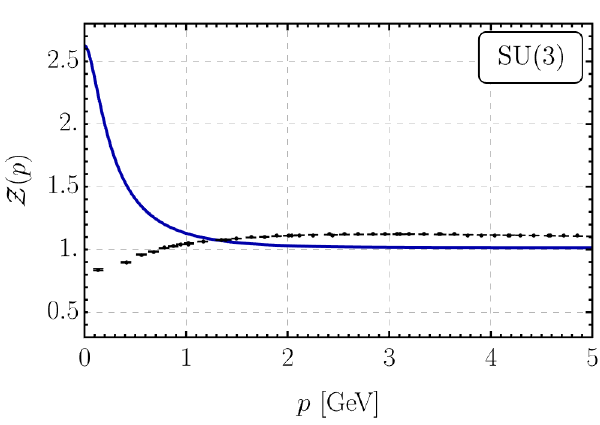}
		\end{minipage} \hfill
		\caption{Quark dressing function $\mathcal{Z}(p)$ with $g_0 \!=\! 7$, $m_\psi \!=\! 37 \, {\rm MeV}$, $\lambda^2 = 2.817$, $m^2 = -0.935$, $M^2 = 7.299$, and $\mathcal{A_Z} \!=\! 1.016$. Lattice data \cite{Oliveira:2018lln} for $SU(3)$ with $N_f \!=\! 2$ and $M_\pi \!=\! 422 \, {\rm MeV}$.}
		\label{Zpsifit}
	\end{figure}

	The optimized choice for $g_0$ and $\mu$, that is, the one that minimizes the $\chi^2$-test, is  $g_0 = 7$ and $\mu = 4 \, {\rm GeV}$, with $m_\psi = 0.037 \, {\rm GeV}$, giving mass-parameters $\lambda^2 = 2.817, m^2 = -0.935, M^2 = 7.299$. 
	The choice $g_0 = 7$ and $\mu = 4 \, {\rm GeV}$ leads to $g(p=\mu) = 3.679$. Therefore, the expansion parameter $N_c\,g^2/(4\pi)^2$ takes the value $0.257$ at the renormalization point.
	The global factors associated with $\mathcal{D}$ and $\mathcal{Z}$ are given by $\mathcal{A_D} = 0.747$ and $\mathcal{A_Z} = 1.016$. The results for $\mathcal{D}(p)$, $\mathcal{M}(p)$ and $\mathcal{Z}(p)$ in comparison with lattice data, adopting the parameters above can be seen in Figs.~\ref{Dpsifit}, \ref{Mpsifit}, and~\ref{Zpsifit}, respectively. 

	In Fig.~\ref{Dpsifit} we observe that, despite the intricate expression for the RGZ gluon propagator at one-loop order, we can choose  mass-parameters allowing for a good agreement with lattice data. The qualitative behavior of the unquenched gluon propagator is very similar to the quenched one, and it also achieves a finite value at $p=0$.
	
	Next to that we consider the quark mass function $\mathcal{M}(p)$, exhibited in Fig.~\ref{Mpsifit}. We emphasize that in this case, we are not fitting any free parameter but rather using the previously fixed  mass-parameters and providing a prediction for given values of $g_0$ and $\mu$. The one-loop result not only captures the correct qualitative behavior, but also gives a reasonably accurate agreement from the quantitative point of view, as one can see in Fig.~\ref{Mpsifit}. We remark that $\mathcal{M}(p)$ achieves a finite value at $p=0$.  There is still room for improvement, since we are not considering the correct RG flows for all the parameters, but only using a toy model to mimic the gauge coupling running~\eqref{RunningCoupling}. However, to perform the complete RG analysis in the RGZ framework using an IR-safe scheme is far beyond the scope of the present work.
	
	Finally, we discuss the quark dressing function $\mathcal{Z}(p)$, showed in Fig.~\ref{Zpsifit}. Our result clearly does not have a good agreement with lattice data. In particular, the deep IR behavior is even qualitatively different. It is worth pointing out, however, that a very similar feature was also observed in \cite{Pelaez:2014mxa}, where the authors realized that this mismatch was due to the  smallness of the one-loop contribution. In this case, the two-loop contributions are not negligible for an accurate description of $\mathcal{Z}(p)$, a fact confirmed later by performing the explicit two-loop computation~\cite{Barrios:2021cks}. Therefore, it is reasonable to expect that the same feature will happen in the RGZ case, that is, the two-loop result will play a pivotal role for an accurate description of the lattice data. 
	
	At this point, it is important to compare our findings  with earlier works regarding the matter sector in the RGZ framework, contextualizing these different approaches. One can propose a non-minimal way of coupling matter fields to the gluons in the RGZ scenario, such that the quark propagator exhibits non-perturbative features already at the tree-level. In \cite{Capri:2014bsa,Dudal:2013vha,Capri:2017abz} the authors introduced the matter fields assuming the presence of a horizon-like term in the matter sector, in analogy with what happens in the gauge sector due to the restriction to the Gribov region. This term is non-local but as in the GZ framework, can be localized by means of auxiliary fields. Furthermore, as it happens within the RGZ setup, the localizing fields acquire their own dynamics and form condensates which shall be added to the local action. The resulting local action is renormalizable \cite{Capri:2014fsa,Capri:2021pye}. The horizon-like term introduced in the matter sector induces new vertices and propagators and, in particular, endows the tree-level matter-field propagator a non-perturbative nature as discussed as pointed out in \cite{Capri:2014bsa,Dudal:2013vha,Capri:2017abz}. It turns out that such propagator fits very well the lattice data in the infrared. The drawback of this non-minimal approach is the inclusion of new massive parameters without a clear geometrical interpretation in contrast to what happens in the gauge sector. The non-trivial term added to the action in the non-minimal approach is given by
	\begin{align}\label{key1}
		S_{\rm M} \!= \! g^2 M^3_{\rm H}\!\! \int \!\! {\rm d}^4x {\rm d}^4y \, \psi^i(x) (T_a)^{ij} \left[\mathcal{M}^{-1}_{xy}\right]^{ab}  (T_b)^{jk} \psi^k(y)\,,
	\end{align}
with ${M}_{\rm H}$ being a massive parameter. The non-locality arises from the inverse of the FP operator $\mathcal{M}$. The localization of \eqref{key1} is achieved by the introduction of two pairs of auxiliary fields $(\bar{\lambda},\lambda)$ and $(\bar{\zeta},\zeta)$ which have anti-commuting and commuting nature, respectively. A condensate of the form. 
	\begin{align}\label{key2}
	S_{\rm M, c} = \mu^2_{\rm H} \int {\rm d}^4x\left(\bar{\lambda}^{ai}\lambda^{ai}+\bar{\zeta}^{ai}\zeta^{ai}\right)\,,
	\end{align}
is added on top of the local version of \eqref{key1}.
Through the inclusion of the above horizon-like term in the matter action, they obtained a fermion propagator at tree-level suitable to fit the lattice data of \cite{Parappilly:2005ei}, with the following expression for the quark mass function,
	\begin{align}\label{key3}
	\mathcal{M}_{\rm nm}(p) = \frac{N_c^2-1}{2N_c}\frac{g^2 M_{\rm H}^3}{p^2 + \mu_{\rm H}^2} + m_{\psi}\,,
	\end{align}	
with $m_\psi$ being the bare quark mass. As it is evident from Eq.\eqref{key3}, the tree-level mass function in the non-minimal prescription already carries a non-trivial momentum dependence. At vanishing momentum, the mass function reaches a non-vanishing value while it decreases for increasing momentum. This is precisly the qualitative behavior one expects for the mass function and what is measured in lattice simulations. There are two new mass-parameters in this prescription, namely, $M_{\rm H}$ and $\mu_{\rm H}$. At tree-level, those parameters are completely independent from the gluonic dynamics and thus free to be adjusted in a fitting procedure for the quark mass function. In principle, including quantum corrections, a mixture between those sectors will show up. Moreover, those parameters could be fixed by their own gap equations, but as we already saw in the case of the mass parameters within the RGZ scenario, this is not useful for the present purposes.
	
Although the mass function displays the correct qualitative behavior already at the tree-level and features free parameters that allow for a fitting with available lattice data, the dressing function $\mathcal{Z}(p)$ is constant (and equal to one) in this approximation. This is not what is reported from lattice simulations, see Fig.~\ref{Zpsifit}, since below $p^2 = 1 \, {\rm GeV}^2$ it decreases. Nevertheless, the deviation is not too strong and for some purposes, such a tree-level result might suffice.
	
The non-minimal scheme presents a reasonable proposal for introducing matter in the RGZ framework. Indeed, this approach has the advantage that the matter fields propagators acquire a non-perturbative behavior already at tree-level, in a similar way as the gauge fields. As such, the quark propagator is suitable to fit the available lattice data quite reasonably. However, the main drawback is that, differently from the gauge sector counterpart, the inclusion of this horizon-like term in the matter sector lacks a clear geometrical motivation and introduced new mass parameters with an unclear dynamical origin. Hence, since the minimal scheme already achieves a good agreement with lattice data, at least for the mass function, which comes out as a prediction under the methodology employed in this work. The success achieved by the CF model in describing the dressing function when higher-orders corrections are taken into account tells us that it is likely that the minimal prescription will be enough to reproduce lattice data and there is no need to introduce extra free parameters.

	\section{Conclusions}\label{SecConclusions}

	In this work, we have computed the quark propagator at one-loop order by introducing fermionic matter minimally coupled with the gluons in the RGZ framework. We compared our findings with available lattice data \cite{Oliveira:2018lln}. Furthermore, we have compute the modification in the RGZ gluon propagator at one-loop order caused by the inclusion of fermionic matter, i.e., the unquenched gluon propagator, also comparing it with lattice data \cite{Sternbeck:2012qs}. Finally, we commented on the results obtained in the CF limit. and also compared our findings with a previous proposal of a non-perturbative coupling of matter in the RGZ framework.
	
	Our result for the unquenched RGZ gluon propagator encoded in the form factor $\mathcal{D}(p)$  is in very good agreement with lattice data \cite{Sternbeck:2012qs}, extending the success already achieved for the quenched case~\cite{deBrito:2024ffa}. The quark mass function $\mathcal{M}(p)$ displays a good agreement with lattice data, not only from the qualitative point of view, but also quantitatively. Moreover, this result can be even improved by considering the correct RG flows for all the couplings and parameters, but this is  left for future investigation. Our results for the quark dressing function $\mathcal{Z}(p)$ does not show a good agreement with lattice data in the IR regime. This discrepancy is due to the unusually small one-loop contribution, as pointed out in \cite{Pelaez:2014mxa}, where the authors also observed this feature in the CF model, closely related to the RGZ. Thus, it follows that the two-loop result is not negligible for the precise description of the quark dressing function (or even to capture its qualitative behavior), something that was confirmed latter by an explicit two-loop computation~\cite{Barrios:2021cks} in the CF model. Thus, it is reasonable to expect that the two-loop contribution will also play a relevant role here, allowing for a good agreement with lattice data. 
	
	The minimal scheme adopted here provided a quark mass function in good agreement with lattice data. Moreover, it allowed us to further investigate the quark dressing function, finding a result that is compatible with other one-loop computation present in the literature~\cite{Pelaez:2014mxa}. Thus, regarding the agreement with lattice data, our one-loop computation shows that the minimal scheme seems to be enough to capture the behavior of the propagators, provided that we take into account the limitations of the approximations employed in this work. This is so because the non-perturbative features coming from the gauge sector are transmitted to the quark sector through quantum fluctuations and seem to be enough to describe its infrared behavior, even if we foresee that a complete RG analysis is necessary to find a better agreement  between $\mathcal{M}(p)$  and lattice data, as well as a two-loop computation seems to be required in order to adjust the deep IR behavior of $\mathcal{Z}(p)$, similarly to what happens in \cite{Barrios:2021cks}. 
	
	Taking the CF-model as benchmark,  the RG flow of the gauge coupling plays a fundamental role in the successful comparison between analytical computations and lattice data for the quark mass function. Thus, a detailed study of the RG flows of the RGZ parameters is a pressing issue that deserves investigation. Moreover, we foresee that a two-loop computation could achieve a good agreement with lattice data for the quark dressing function in the same way as it was shown in \cite{Barrios:2021cks}. Such an endeavor in the RGZ scenario constitutes a relevant future investigation, although technically it is significantly more challenging.  An important continuation of  the present research program consists in studying the gauge-matter vertices at one-loop in its fermionic and scalar versions. This is under investigation. Furthermore, a dedicated exploration of the inclusion of dynamical chiral-symmetry breaking effects is essential. Finally, it is worth pursuing the computation of quantum corrections in the context of the non-minimal coupling to matter. Due to the amount of extra propagators and vertices, this is a highly non-trivial task, but could bring very important insights on the nature of the matter coupling to non-Abelian gauge theories in the strongly correlated regime.

	\begin{acknowledgments}
		We are grateful to Orlando Oliveira and Andre Sternbeck (on behalf of their collaborators) for providing the lattice data used for the propagators. We are indebted with Marcela Peláez and Nahuel Barrios for valuable technical discussions that helped us to improve our results, and also to Letícia Palhares and Bruno Mintz for valuable comments. ADP acknowledges CNPq under the grant PQ-2 (312211/2022-8), FAPERJ under the ``Jovem Cientista do Nosso Estado” program (E-26/205.924/2022) for financial support. GPB was supported by the research grant (29405) from VILLUM fonden during a part of this project. PDF acknowledges FAPERJ for financial support under Contract No. SEI-260003/000133/2024 and is grateful for the hospitality at Unesp-Guaratinguetá during part of this work.
	\end{acknowledgments}

	\appendix
	
	\section{Analytical expression for the quark propagator at one-loop}\label{AppAnalExp}

	The RGZ gluon propagator form factor in the Landau gauge at tree-level is given by Eq.~\eqref{GluonRGZtree}. It can be rewritten in a more convenient form as
	\begin{align}
		\mathcal{D}_0(p) = \frac{R_1}{p^2 + \mu_1^2}  + \frac{R_2}{p^2 + \mu_2^2},
	\end{align}
	where we define
	\begin{align}
		R_1 &= \frac{1}{2} \left(1 + \frac{m^2 - M^2}{\Omega}\right), \nonumber \\
		R_2 &= \frac{1}{2} \left(1 + \frac{M^2 - m^2}{\Omega}\right),
	\end{align}
	and
	\begin{align}
		\mu_1^2 &= \frac{1}{2}\left(m^2 + M^2 + \Omega\right), \nonumber \\
		\mu_2^2 &= \frac{1}{2}\left(m^2 + M^2 - \Omega\right),
	\end{align}
	with the definition $\Omega = \sqrt{(m^2-M^2)^2 - 8 \gamma^4 g^2N}$.
	
	The divergent part of the quantum correction  $\delta\Gamma_{\psi \bar{\psi}}(p)$ at one-loop order (adopting $d = 4 - 2 \epsilon$) is given by:
	\begin{align}\label{key}
		\Gamma_{\psi }\vert_\text{div} = -\frac{3 g^2 C_F  }{16 \pi^2 \varepsilon} (R_1 + R_2) m_\psi,
	\end{align}
	where we are using $C_F \equiv \frac{N_c^2 -1}{2 N_c}$.
	The finite part can be written as $ \Gamma^{(A)}_\psi + \Gamma^{(B)}_\psi$, where:
	\begin{align}\label{key}
		\Gamma^{(A)}_\psi &=  i \slashed{p} \frac{ g^2 C_F}{16 \pi^2}  \Bigg\{(R_1 + R_2)  \nonumber \\
		& -\int_0^1 {\rm d}x \left(3x - 1\right) \left(R_1 \log\frac{\Delta_1}{\mu^2} + R_2 \log\frac{\Delta_2}{\mu^2}\right) \nonumber \\
		&-\int_0^1 {\rm d}x \int_0^{1-x} dy \left[2 p^2 x^2 \left(\frac{R_1}{\tilde{\Delta}_1} +  \frac{R_2}{\tilde{\Delta}_2}\right) \right.  \nonumber \\ 
		& \left. - \left(R_1 \log\frac{\tilde{\Delta}_1}{\mu^2} + R_2 \log\frac{\tilde{\Delta}_2}{\mu^2}\right)\right] \Bigg\} \\
		\Gamma^{(B)}_\psi &= \frac{ 3 g^2 C_F}{16 \pi^2} m_\psi \left[(R_1 + R_2) \left( \gamma -\log 4\pi + \frac{2}{3}\right) \right. \nonumber \\
		&\left. + \int_0^1 {\rm d}x \left(R_1 \log\frac{\Delta_1}{\mu^2} + R_2 \log\frac{\Delta_2}{\mu^2}\right)\right],
	\end{align}
	where we have $\Delta_i = p^2 x \left(1-x \right) + \left( m_\psi^2 - \mu_i^2 \right) x + \mu^2_i$ and $\tilde{\Delta}_i =  p^2 x \left(1-x \right) +  m_\psi^2 x + \mu_i^2 y$. Notice that there are no divergences in the $\slashed{p}$ term.


\bibliography{refs} 


\end{document}